\documentclass[12pt]{article}
\usepackage{cite}
\usepackage{mathptmx}
\usepackage{setspace}

\usepackage{changepage}
\usepackage{dcolumn}
\usepackage[table]{xcolor}
\usepackage{floatrow}
\usepackage{floatrow}   
\usepackage{subcaption}
\usepackage{graphicx,xcolor} 
\usepackage{url}
\usepackage{authblk}
\usepackage{hyperref}

\usepackage[margin=1in]{geometry}
\usepackage{amsmath}
\usepackage{amssymb}
\usepackage{pgf}
\usepackage{comment}

\newtheorem{theorem1}{Theorem}

\definecolor{mynavy}{HTML}{000080}
\definecolor{darkred}{HTML}{8B0000}
\definecolor{mygreen}{HTML}{006400}
\definecolor{mygold}{HTML}{B8860B}

\newcolumntype{d}[1]{D..{#1}}

\title{On the structure of the world economy: An absorbing Markov chain approach}
\author{Olivera Kostoska$^{1,4}$, Viktor Stojkoski$^{2,4}$ and Ljupco Kocarev$^{3,4}$ }
\affil{$^{1}$ Faculty of Economics - Prilep, ``St. Kliment Ohridski'' University, {Bitola, 7000}, North Macedonia \\ 
$^{2}$ Faculty of Economics, Ss. Cyril and Methodius University, {Skopje, 1000}, North Macedonia \\
$^{3}$ Faculty of Computer Science and Engineering, Ss. Cyril and Methodius University, {Skopje, 1000}, North Macedonia \\
$^{4}$ Macedonian Academy of Sciences and Arts, {Skopje, 1000}, North Macedonia  }

\date{\today}

\begin{document}

\maketitle

\begin{abstract}
The expansion of global production networks has raised many important questions about the interdependence among countries and how future changes in the world economy are likely to affect the countries' positioning in global value chains. We are approaching the structure and lengths of value chains from a completely different perspective than has been available so far. By assigning a random endogenous variable to a network linkage representing the number of intermediate sales/purchases before absorption (final use or value added), the discrete-time absorbing Markov chains proposed here shed new light on the world input/output networks. The variance of this variable can help assess the risk when shaping the chain length and optimize the level of production. Contrary to what might be expected simply on the basis of comparative advantage, the results reveal that both the input and output chains exhibit the same quasi-stationary product distribution. Put differently, the expected proportion of time spent in a state before absorption is invariant to changes of the network type. Finally, the several global metrics proposed here, including the probability distribution of global value added/final output, provide guidance for policy makers when estimating the resilience of world trading system and forecasting the macroeconomic developments.
\end{abstract}

\section{Introduction}

The ideas of networked economy, pervasive transmission channels, systemic risk and complexity have become increasingly important after the 2008 financial crisis, but nowadays they are major concern on the impact of global trade tensions. Over the last decades, international fragmentation of production has made a huge transformation in geography and dynamics of international trade. Such fragmentation of production activities has given rise to the Global Value Chains (GVCs) and greatly contributed to reinforce the structural interdependence worldwide. The global production networks are very complex, with flows of value-added representing a final outcome of the complex linkages that exist between firms in different industries and countries over time. The evaluation of these linkages calls for developing new tools that go well beyond the appraisal of bilateral gross trade flows. The network analysis and related metrics are extremely important in assessing the complexity of the whole structure of interactions (direct and indirect linkages) in the world economy, whilst the current research is still in infancy.

So far, various macroeconomic models have been developed, ranging from dynamic stochastic general equilibrium modeling to agent-based macroeconomic models. The former assumes, in a standard setting, an economy that is populated by both an infinitely-lived representative household and a representative firm, with homogenous production technology that is hit by exogenous shocks. The latter, on the other hand, considers an economy populated by heterogenous agents, whose far from equilibrium interactions continuously change the structure of the system. Most of the models are, to some extent, descended from the Leontief’s work on input-output tables~\cite{Leontief-1951}, in which firm/agent interactions are mainly characterized by the global production networks. The latter are typically described with the so-called multi-regional input-output (MRIO) models that combine, in a coherent framework, national input-output and trade flow tables. In addition to tracking GVCs, together with other methodologies, such as, for example, life cycle assessment, material flow analysis, MRIO models have been used for sustainability analysis addressing a wide range of policy and research questions regarding the impacts of global production networks \cite{Peters8903}. There are several independently constructed global MRIO databases. In this paper, all theoretical results are accompanied with numerical computations of the World Input-Output Database (WIOD) \cite{timmer2015illustrated}.

Basically, our approach is motivated by the recent observations in two strands of theoretical and empirical literature: shock propagation in economic networks and value chain positioning. 
The production network can work as a channel for propagating shocks throughout the economy. The possibility that substantial aggregate fluctuations may originate from microeconomic shocks has long been abandoned in the literature (see for example \cite{lucas1977understanding}). This is mainly due to the `diversification argument', which states that, in an economy consisting of $n$ industries hit by independent shocks, aggregate fluctuations would have a magnitude that is roughly proportional to $1/\sqrt{n}$, a minor effect at high levels of disaggregation. This argument, however, disregard the input-output linkages between different firms and industries operating as a propagation channel of idiosyncratic microeconomic shocks throughout the economy~\cite{Acemoglu-2012,carvalho2018production,baqaee2019macroeconomic,carvalho2014micro,barrot2016input,acemoglu2016networks}. The other main question of interest is the representation of network-originated macro fluctuations in terms of the economy’s structural parameters. In line with the key observations of~\cite{Acemoglu-2012}, different roles various sectors play as input suppliers to others may generate sizable aggregate volatility when compared against the standard diversification argument rate. Microeconomic shocks may propagate over the network, but if propagated symmetrically, they would average out, and thus, would have minimal aggregate effects (hence the diversification argument remains applicable). The diversification argument would not hold, however, if intersectoral input-output linkages do not display such symmetries. Put differently, when sectors are highly asymmetric as input suppliers, even with a large number of industries, shocks to sectors that are more important suppliers propagate strongly to the rest of the economy producing significant aggregate fluctuations. Similarly, an industry will take a more `central' position in the network if it plays a more important role as an input supplier to other central industries, and thus, it will be more influential in determining the aggregate output (see for example the Bonacich centrality). This statement goes in line with the intuition that productivity shocks to an industry with more direct or indirect downstream customers should have sizable aggregate effects.

The second strand of the literature pays particular attention to quantifying the relative production line position of industries/countries (or country-industry pairs) in (global) value chains. By dramatically changing the organization of world production, the rise of GVCs has laid in focus the specialization of countries within the global value chains. In point of fact, if shocks propagate downstream (to customer industries, supply-side shocks) or upstream (to input-supplying industries, demand-side shocks), the economic condition of a certain industry/country/country-industry pair is largely dependent on its relative position along the global output supply chain and the global input demand chain. 
The positions are usually measured relative to final consumers and primary inputs as downstream and upstream ends of value chains.  
The ``upstreamness from final consumption'' and ``downstreamness from primary factors'' are two numerical estimates (based on the length of output supply/input demand chains, respectively) that measure the country`s/industry`s/country-industry pair`s position in global production processes \cite{antras2013organizing,miller2017output,fally2012production,antras2012measuring}. However, two limitations, as argued by \cite{wang2017characterizing} may possibly reduce the reliability of ‘upstreamness’ and ‘downstreamness’: first, they all begin with an industry’s gross output, whilst the production chain should begin with the industry’s primary inputs (or the sources of value added). Second, these measures do not imply each other and might suggest inconsistent positions for the same country-industry pairs.

This paper, by proposing two (input and output) discrete-time absorbing Markov chains, opts to investigate the positions, length of chains and structural interdependence of the world economy (especially in regard to propagation and effects of economic shocks) from a completely different perspective. The first theoretical linkage between Markov chain and Leontief input-output models was derived by Solow in 1952 \cite{Solow-1952}. In spite of the great variety of mathematical properties, so far, however, world economic networks have not been fully examined with Markov chain formalism. Here, we introduce a random endogenous variable pointing to the number of intermediate sales/purchases before absorption (final demand/primary inputs or sources of value added). The variance of this variable can serve as a useful guide for policy makers when determining the “optimal level of fragmentation”. Further, our results reveal the same quasi-stationary product distribution of both chains, that is, the product distribution does not depend on the type of the network (world input or world output network). Finally, our proposed global metrics (Global Domestic Value-Added, Global Import-Export Value-Added, Global Domestic Final Use and Global Import-Export Final Use) and our measures about the probability distribution of value added (global value added matrix) or final output (global final demand matrix) across countries allow for a more precise determination of the global production chain lengths than any presently available.
The rest of the paper is organized as follows. Section \ref{sect-methods} describes context of the paper and data collection instruments, and outlines the major analysis methods. Section \ref{sect-results} presents the main findings and discusses results with reference to preceding research, as well as to their practical and policy implications. The last section provides the main conclusions of the research.

\section{Materials and Methods} \label{sect-methods}

\subsection{World Input-Output Database} 

The analyses performed here are based on data from 2016 release of the World Input-Output Database (WIOD).  The latter contains annual time-series of world input-output tables. A World Input-Output Table (WIOT) can be viewed as a set of national input-output tables that are linked together through detailed bilateral international trade statistics. In short, WIOT provides a comprehensive summary of all international transactions between industries and final consumers.
%
%
%
%
The columns in the WIOT contain information about production processes. 
When expressed as ratios to gross output, the cells in a column deliver information on the shares of inputs in total costs. Such a vector of cost shares in gross output is commonly referred to as a production technology.
Products can be purchased by final users, or final demand expenditures (households consumption, government consumption, gross fixed capital formation, and change in inventories) or integrated into other goods and services (intermediates).  The distribution of the output of industries over user categories is indicated in the rows of the table. An important accounting identity in the WIOT is that gross output of each industry (given in the last element of each column) is equal to the sum of all uses of the output from that industry (given in the last element of each row).

The 2016 release covers forty-three countries, including 28 EU member states and 15 other major economies, for the period from 2000 to 2014. The countries are selected on the basis of both data quality and aspirations to include a major part of the world economy. 
These countries together cover more than 85\% of the world GDP; a model is estimated for the remaining non-covered part of the world economy, or the `rest of the world' region.
The WIOD is structured according to the recent industry and product classification ISIC Rev. 4 (or equivalently NACE Rev. 2), with the underlying WIOTs covering 56 industries. Further information about the included countries and industries can be found in~\cite{timmer2015illustrated,timmer2016anatomy}.  The dataset is available at \url{http://www.wiod.org/home}.

As the data has more than 2 dimensions, it should be arranged as multidimensional arrays, which are often called tensors.  
The order of a tensor is the number of dimensions. Vectors (tensors of order one) are denoted by boldface lowercase letters, e.g., $\mathbf{a}$. Matrices (tensors of order two) are indicated by boldface capital letters, e.g., $\mathbf{A}$. Higher-order tensors (order three or higher) are symbolized   by math calligraphy letters, e.g., $\mathcal{X}$. Scalars are designated by lowercase or  uppercase letters, e.g., $a$ or $A$.
The $i$-th entry of a vector $\mathbf{a}$ is denoted by $a_i$, element $(i, r)$ of a matrix $\mathbf{A}$  is denoted by $a_{ir}$ or $a_i^r$, element $(i, j, r)$ of a third-order tensor $\mathcal{X}$ is indicated by $x_{ijr}$ or $x_{ij}^r$, and element $(i,j,r,s)$ of a forth-order tensor $\mathcal{Z}$ is represented by $z_{ijrs}$ ($Z_{ijrs}$) or $z_{ij}^{rs}$  ($Z_{ij}^{rs}$).

The WIOD includes detailed data for $J$ countries (indexed by $\hat{i}$ or $\hat{j}$) and $S$ sectors (indexed by $r$ or $s$) organized as two tensors:   4-order tensor  $\mathcal{Z} \in \mathbb{R}^{J\times J\times S\times S}$ with entries $z_{\hat{i} \hat{j}}^{rs}$ describing the intermediate purchases (input flows) by industry $s$ in country $\hat{j}$ from sector $r$ in country $\hat{i}$; and  3-order tensor $\mathcal{F} \in \mathbb{R}^{J\times J\times S }$ with entries $f_{\hat{i} \hat{j}}^r$ denoting the final use in each country $\hat{j}$ of output originating from sector $r$ in country $\hat{i}$.
In addition, the WIOD includes matrices $\mathbf{F}$,  $\mathbf{X}$,  and $\mathbf{W}$, uniquely  determined by the tensors $\mathcal{Z}$ and $\mathcal{F}$. The entries of these matrices are given as follows:
\begin{align}
f_{\hat{i}}^r &= \sum_{\hat{j} =1 }^J f_{ \hat{i} \hat{j}}^r, \label{eq-matrix-F-1} \\
x_{\hat{i}}^r &= \sum_{s=1}^S \sum_{\hat{j}=1}^J z_{\hat{i} \hat{j}}^{rs} + 
f_{\hat{i}}^r, \label{eq-matrix-X-1} \\
%
%
w_{\hat{j}}^s  &= x_{ \hat{j}}^s - \sum_{r=1}^S \sum_{ \hat{i}=1}^J z_{\hat{i} \hat{j}}^{rs}.  \label{eq-matrix-V-1}
\end{align} 
%
%
where $x_{\hat{i}}^r$ represents the value of gross output originating from sector $r$ in country $\hat{i}$;  the element $f^r_{\hat{i}}$  stands for the value of output from sector $r$ in country $\hat{i}$ intended for final consumers worldwide; and $w^s_{\hat{j}}$ indicates the country’s  $\hat{j}$ value-added employed in the production of an industry $s$.

An $N$ th-order tensor is an element of the tensor product of $N$ vector spaces, each of which has its own coordinate system. Slices are two-dimensional sections of a tensor, defined by fixing all but two indices. 
Matricization, also known as unfolding or flattening, is the process of reordering the elements of an $N$-th order array into a matrix. For instance, a $2 \times 2 \times 3 \times 3$ tensor can be arranged as a $6 \times 6$ matrix or a $2 \times 18$ matrix, and so on. It is also possible to vectorize a tensor; for example, $2 \times 2 \times 3 \times 3$ tensor can be arranged as a 36 dimensional vector. 

The world input-output table is obtained by unfolding the tensors $\mathcal{Z}$ and $\mathcal{F}$ into $JS \times JS$ and $JS \times S$ matrices respectively, and by unfolding the matrices $\mathbf{F}$, $\mathbf{X}$, and $\mathbf{W}$ into $JS$ vectors. Therefore, WIOT consists of the following elements:
\begin{itemize}

\item $n \times n$ matrix $\mathbf{Z}= [ z_{ij}]$ with $i = (\hat{i},r)$ and $j = (\hat{j},s)$  and $n = JS$, so that $ij$ element of the matrix describes the sales of intermediates from country-industry pair $i$ to country-industry pair $j$
\begin{align} \label{eq-wiot-matrix-Z}
\mathbf{Z} &= \begin{bmatrix}
z_{11} & \ldots & z_{1n} \\
\vdots & \ddots & \vdots \\
z_{n1} & \ldots & z_{nn} \\
\end{bmatrix}.
\end{align}

\item $n$ dimensional final demand  vector $\mathbf{f} = \left[ f_1, \ldots, f_n \right]^T$ with the $i$-th entry describing the sales from country-industry pair $i$ to final users.

\item  $n$ dimensional gross-output vector $\mathbf{x} = \left[ x_1, \ldots, x_n  \right]^T,$ where $ x_i = \sum_{j} z_{ij} + f_i. $

\item  $n$ dimensional value-added vector $\mathbf{w} = \left[ w_1, \ldots, w_n  \right]^T,$ where $w_i = x_i - \sum_{j} z_{ji}.$
\end{itemize}

\subsection{World-input and world-output networks}

Consider a world economy with $J \geq 1$ countries (economies):   country-1, \ldots, country-$J$ and $S \geq 1$ sectors (industries): sector-1, \ldots, sector-$S$  as a network $G= (V, E)$ of $n=JS$ nodes in which each node represents a country-industry pair, where $V = \{ 1, \ldots, n\}$ is the set of nodes and $E$ is the set of edges to be defined shortly.  Country-industry pairs $(\hat{i},r)$ are mapped to the nodes in $V$ with $(\hat{i},r) \rightarrow (\hat{i}-1)S + r $, for $ \hat{i}=1, \ldots, J$ and $r=1, \ldots, S$. 
Note that the nodes $1 , \ldots, S$ correspond to the country-1, the nodes $S+1, \ldots, 2S$ are related to the country-2, and so on.

Let us first define  $i = (\hat{i},r)$ and $j= ( \hat{j},s )$,  so that the  country-industry pairs are indexed by $i$ and $j$, $i,j = 1, \ldots, n$. Next, we associate two networks (world-input network and world-output network) with the vertex set $V = \{1, 2, \ldots, n  \}$.  
\textit{World-input network} is represented by the adjacency input matrix $\mathbf{A} = [a_{ij}]$  for which $a_{ij} \equiv z_{ij}/x_j$. This normalization will be called  ``world-input'' network, since, along the output supply chain, the country-industry pair $i$ sells intermediate inputs to other country-industry pairs $j$’s in the world economy (the corresponding links are denoted by the input coefficients $a_{ij}$).

For the \textit{world-output network},  $ \mathbf{B}=[b_{ji}]$ represents an adjacency output matrix of the second normalization with a typical element $b_{ji} \equiv z_{ji}/x_j$. 
This specification will be called (is known as) ``world-output'' network, since, along the input demand chain, the country-industry $i$ buys intermediate inputs worldwide (the corresponding links from all country-industry pairs $j$’s to $i$ are denoted by the output coefficients $b_{ji}$). 
Note that the matrices $\mathbf{A}$ and $\mathbf{B}$ are similar, ($\mathbf{B} = \mathbf{X}_{dg}^{-1} \mathbf{A} \mathbf{X}_{dg}$, where $\mathbf{X}_{dg}$ is the diagonal matrix with elements of $\mathbf{x} = [x_1, \ldots, x_n]^T$ along its diagonal and zeros otherwise),  share the same eigenvalues, and their largest eigenvalue, $\lambda$, is real with  $\lambda < 1$. 

The world production here is modelled linearly, that is $(\mathbf{A}, \mathbf{f})$ or  $(\mathbf{B}, \mathbf{w})$, where $\mathbf{f} = [ f_1, \ldots, f_n]^T$ and $\mathbf{w} = [ w_1, \ldots w_n ]^T$. 
Let $\mathcal{L} = ( \mathbf{I} - \mathbf{A})^{-1}$ and $\mathcal{G} = ( \mathbf{I} - \mathbf{B})^{-1} $ be the Leontief-inverse matrix and Ghosh-inverse matrix, respectively. Then Eq.~(\ref{eq-matrix-X-1}) and  Eq.~(\ref{eq-matrix-V-1}) can be rewritten in a compact form as~\cite{antras2013organizing,miller2017output},
\begin{eqnarray}
\mathbf{x} & = & \mathcal{L} \, \mathbf{f} \label{eq-leontief} \\
\mathbf{x} & = & \mathcal{G}^T \mathbf{w} \label{eq-ghosh} 
\end{eqnarray}
The two well-established metrics in input-output economics indicating the country-industry’s (weighted) average position in global value chains \cite{antras2013organizing,miller2017output}, the output upstreamness (or upstreamness) (OU), $\mathbf{u} = [u_1, \ldots, u_n]^T$, and the input downstreamness (or downstreamness) (ID), $\mathbf{d} = [d_1, \ldots, d_n]^T$, are defined as $\mathbf{u} = \mathcal{G} \, \mathbf{1} $ and $\mathbf{d} = \mathcal{L}^T \mathbf{1}$, where $\mathbf{1}$ is a length-$n$ column vector whose entries are all 1. 
In world-output network, the country-industry pairs with large values of $d_i$ will produce complex and strong intermediate input demand edges with similar pairs (and vice versa for small values). In world-input network, on the other hand, the country-industry pairs with large $u_i$ values will produce complex and strong intermediate output supply edges with similar pairs (and vice versa for small values).


\subsection{Absorbing Markov chains}

We next turn to proposing alternative measures of the country-industry’s average position in global value chains. These measures can be estimated for both the output supply chain (hence relative to final consumption) and the input demand chain (thus relative to primary inputs). 
In order to understand the structure and organization of the world production, the input and output networks here are associated with two homogeneous discrete-time absorbing Markov chains, that is the \textit{input-chain} and the \textit{output-chain}. The state space of both chains is $V \cup \{ 0 \} = \{0, 1, 2, \ldots, n \}$, while the vertex set $V$ is considered as a set of transient states. For the world-input network,  the absorbing state  $0$ represents final use of output. On the other hand, for world-output network,  the absorbing state  $0$ represents primary factors of production (or sources of value added).  
If we define   
\begin{eqnarray}
\gamma_i & = & \frac{f_i}{x_i}  \label{eq-gamma-data} \\
\delta_i & = & \frac{w_i}{x_i}  \label{eq-delta-data} 
\end{eqnarray}
the transition matrix $\mathbf{P}_{in}$ of the absorbing Markov input-chain reads 
\begin{eqnarray}  \label{eq-matrix-Pin}
{\mathbf{P}}_{in} &=& 
\begin{bmatrix}
1 & \mathbf{0}^T \\
 \boldsymbol{\delta}  & \mathbf{A}^T 
\end{bmatrix},
\end{eqnarray}
while the transition matrix $\mathbf{P}_{out}$ of the absorbing Markov output-chain is  
\begin{eqnarray} \label{eq-matrix-Pout}
{\mathbf{P}_{out}} &=& 
\begin{bmatrix}
1 & \mathbf{0}^T \\
\boldsymbol{\gamma} & \mathbf{B} 
\end{bmatrix},
\end{eqnarray}
where $T$ denotes transpose operator, $\boldsymbol{\gamma} = [\gamma_1, \ldots, \gamma_n]^T$, 
$\boldsymbol{\delta} = [\delta_1, \ldots, \delta_n]^T$, and 
$\mathbf{0}$ is a length-$n$ column vector whose entries are all 0. 
Both matrices $\mathbf{P}_{in}$ and $\mathbf{P}_{out}$ are row stochastic.  The column stochastic matrix  $\mathbf{A}$  and row stochastic matrix $\mathbf{B} $ are the one-step transition probabilities of the (sub)Markov chain on $V$  and $\delta_i$ and $\gamma_i$ are the one-step transition probabilities of absorption into the state $0$.  

Let us analyse the world input/output network with a Markov process  (input-chain or output-chain).  Assume that the process starts in state $i \in V$ at time 0, and let $Y_{ij}^{(t)}= 1$ (or 0) if the
process is (or is not) in the state $j$ at time $t$. 
Further, let $X_{ij} = \sum_{t=0}^\infty Y_{ij}^{(t)}$ be a random variable representing the number of visits to state $j$ before absorption and $X_i = \sum_{j \in V} X_{ij}$ be another random variable -- the time to absorption.
To simplify notation, we write $\mathbf{P}$ for both matrices  $\mathbf{P}_{in}$ and $\mathbf{P}_{out}$ and allow for $\mathbf{Q}$ to be  either $\mathbf{A}^T$  or $\mathbf{B}$. 
%
%
%
From the standard theory of absorbing Markov chains, the following equations can be derived \cite{kemenyfinite}:
\begin{eqnarray}
\mathbf{L} \equiv \left[ \mathbb{E} [ X_{ij} ] \right] & = & \left(\mathbf{I} - \mathbf{Q}\right)^{-1}  \label{eq-mean-Q} \\
\mathbf{L}_2 \equiv \left[\mbox{Var} [ X_{ij} ] \right] & = &  \mathbf{L} ( 2 \mathbf{L}_{dg} - \mathbf{I}) - \mathbf{L}_{sq}  \label{eq-var-Q} \\
\mathbf{g} \equiv \mathbf{g}(\mathbf{Q}) \equiv \left[  \mathbb{E} [X_1], \ldots, \mathbb{E} [X_n] \right]^T& = & \mathbf{L} \mathbf{1}  \label{eq-mean-X} \\
 \mathbf{h} \equiv \mathbf{h}(\mathbf{Q}) \equiv  \left[ \mbox{Var} [X_1], \ldots, \mbox{Var} [X_n]\right]^T & = & (2 \mathbf{L} - \mathbf{I} ) \mathbf{g} - \mathbf{g}_{sq}
\label{eq-var-X}
\end{eqnarray}
where $\mathbb{E}[X]$ and $\mbox{Var}[X]$ are expectation and variance of the random variable $X$, respectively, 
$ \mathbf{L}_{dg} = [ \ell_{ii} ]$ is a diagonal matrix, $ \mathbf{L}_{sq} = [ \ell_{ij}^2 ]$,    
$\mathbf{1}$ is a length-$n$ column vector whose entries are all 1, and 
 $ \mathbf{g}_{sq} = [ g_1^2, \ldots, g_n^2 ]^T$.

When $\mathbf{Q} = \mathbf{A}^T$, the transpose of the matrix defined with Eq.~(\ref{eq-mean-Q}) coincides with the Leontief-inverse matrix $\mathcal{L}$. For $\mathbf{Q} = \mathbf{B}$, Eq,~(\ref{eq-mean-Q}) reduces to the Ghosh-inverse matrix  $\mathcal{G}$.  
Moreover, from Eq.~(\ref{eq-mean-X}), it follows that the expected number of steps before absorption is characterized with vectors (for the output and input networks) $\mathbf{u}$ and $\mathbf{d}$. The metric $\mathbf{u}$ is a measure of distance of a country-industry pair from the final demand. 
Therefore,  $u_i$ describes ``how far'' (in expected number of steps) the production of a county-industry pair is from the final use (or ``average production line position''). 
The second quantity $\mathbf{d}$  is a measure of distance of a country-industry pair from the primary factors of production (or sources of value-added). In another words, $d_i$ measures ``average distance from primary inputs suppliers''. 
Not all products, however, need to have their production split into multiple stages. Services, for example, are less inclined to vertical specialization when supplier is required to have a close contact with the consumer. The variance (of the number of steps before absorption) can help measure the volatility (or the risk) a country-industry pair assumes when determining the production chain lengths. It could therefore permit the country-industry pairs to develop better models of production by optimizing the level of fragmentation (that is a trade-off between higher transaction/coordination costs and lower costs of production).

For both absorbing Markov chains, the input-chain and the output-chain, the set of transient state $V$ is irreducible. In terms of the  world-input  network, it means that for arbitrary two country-industry pairs $i$ and $j$, even for those directly not connected ($a_{ij}=0$), the country-industry pair $i$, after finite number of jumps (hops/steps), sells intermediate inputs to the  country-industry pair $j$. 
Similarly, irreducibility of the world-output network implies that an arbitrary country-industry pair, after finite number of purchases, buys intermediate inputs from all other country-industry pairs worldwide. 
However, since the Markov chains are absorbing, eventually, intermediate output sales and input purchases reach the absorbing state –- state from which further jumps are impossible (e.g. final consumers in the  world-input network and primary factors or sources of value added).


\begin{figure}[t!]
\centering
\includegraphics[width=\linewidth]{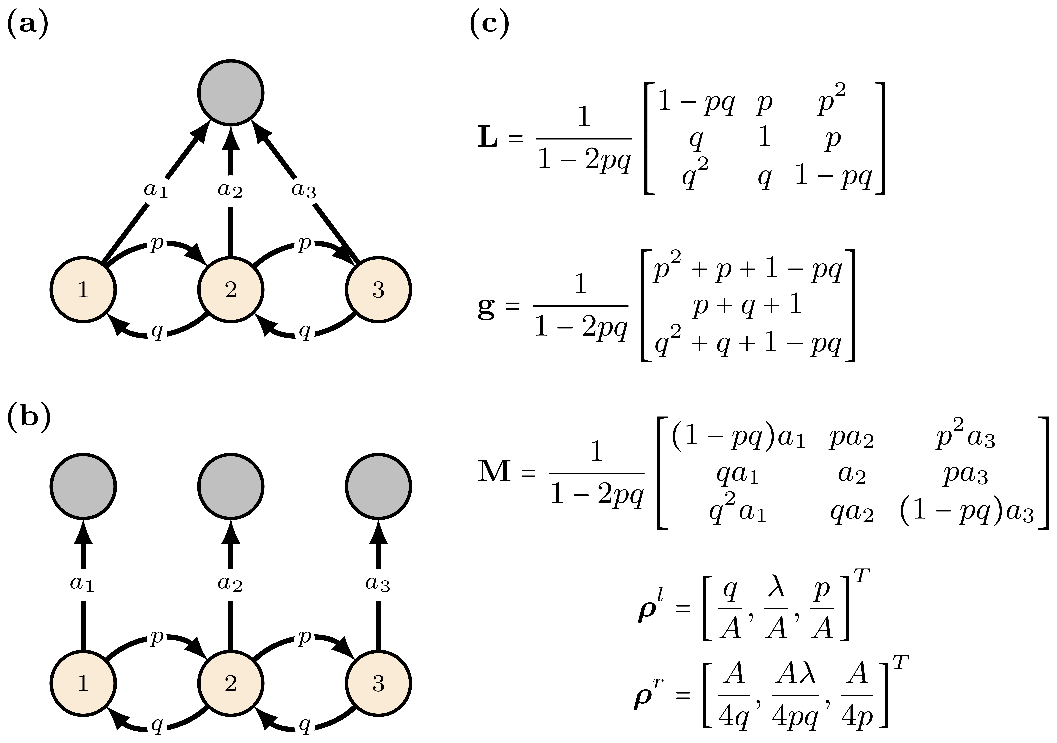}
\caption{A simple example with three countries $C_1, C_2, C_3$ and one industry $I$. The world economy includes three pairs: $1 \equiv (C_1,I)$, $2 \equiv (C_2,I)$, and $3 \equiv (C_3,I)$.
The first pair produces intermediate products in amount of $p$ percentage of the total output for the second pair, and final products in amount of $a_1$ percentage of the total output for the country $C_1$. 
The second pair produces intermediate products in amount of $q$ percentage of the total output for the first pair, and intermediate products in amount of $p$ percentage of the total output for the third pair, and  final products in amount of $a_2$ percentage of the total output for the country $C_2$. 
The third pair produces intermediate products in amount of $q$ percentage of the total output for the second pair, and final products in amount of $a_3$ percentage of the total output for the country $C_3$.
(a) The world economy is represented as an absorbing Markov chain with one absorbing state. In this case $\boldsymbol{\gamma} = [a_1, a_2, a_3]^T$.  
(b) The world economy is represented as an absorbing Markov chain with three  absorbing states. In this case, $\mathbf{D}_{\eta}$, Eq.~(\ref{eq-matrix-Pout-1}), is a $3 \times 3$ diagonal matrix with elements $a_1$, $a_2$, and $a_3$.  
(c) The Ghosh-inverse matrix $\mathbf{L} = ( \mathbf{I} - \mathbf{B}  )^{-1} = \mathcal{G}$, Eq.~(\ref{eq-mean-Q}), the output upstreamness, $\mathbf{g}$, Eq.~(\ref{eq-mean-X}), and the matrix $\mathbf{M}$. The largest eigenvalue of the matrix $\mathbf{B}$ is $\lambda = \sqrt{2pq}$.
The left and right eigenvectors are also shown with $A = p+q + \lambda$. For this example, the product distribution does not depend on $p$ and $q$ and is equal to $\{1/4, 1/2, 1/4   \}$.
}
\label{fig:example-1}
\end{figure}

Next, we consider two quasi-stationary distributions which are derived, roughly speaking, from observing only those realizations in which the time to absorption is long.
Assume that the process starts in state $i$ with probability $\pi_i$. 
Let $\mathbf{P}^{t} = [ p_{ij}^{(t)} ]$. The probability that the process has
been absorbed by time $t$ is given as $ \sum_{i \in V} \pi_i p^{(t)}_{i0}$. 
Provided that the process is not absorbed at time $t$, the conditional probability that the process is in state $j$ at time $t$ is computed as follows~\cite{darroch1965quasi}: 
\begin{eqnarray} \label{eq-q-distr-1}
\mbox{Pr [in state $j$ at time $t$ | not absorbed by time $t$] } &= &  \nonumber  \\ 
\frac{\sum_{i \in V} \pi_i p_{ij}^{(t)}  }{ \sum_{i\in V} \pi_i \left[ 1 - p_{i0}^{(t)}  \right]  } =   \rho_j^l + O \left(  \left( \frac{| \lambda_2 |}{\lambda}  \right)^t   \right) 
\end{eqnarray}
where $\boldsymbol{\rho}^l = [\rho_1^l, \ldots, \rho_n^l     ]^T$ is the left dominant eigenvector of the matrix $\mathbf{Q}$, $\lambda_2$ is the second largest eigenvalue of the matrix $\mathbf{Q}$, and we have assumed, for simplicity only, that its multiplicity is 1.  This vector is normalized, $\sum_i \rho_i^l =1$, so it represents a quasi-stationary distribution of the absorbing Markov chain. 
In other words, the quasi-stationary distribution represents the proportion of time the process spends in a transient state, that is the number of times a particular country-industry pair contributes to its own production stages, or becomes an intermediate in its own supply/demand chains before absorption (given that the time to absorption is long).
Note that the limit as $t \to \infty$ is $ \rho_j^l$  which is independent of the probability distribution $\boldsymbol{\pi} = [\pi_1, \ldots, \pi_n  ]^T$.
Equation (\ref{eq-q-distr-1}) can be further generalized as \cite{darroch1965quasi}: 
\begin{eqnarray} \label{eq-q-distr-2}
\mbox{Pr [in state $j$ at time $\tau$ | not absorbed by time $t$] } &= &  \nonumber  \\ 
 \rho_j^l \rho_j^r + O \left(  \left( \frac{| \lambda_2 |}{\lambda}  \right)^\tau   \right) + O \left(  \left( \frac{| \lambda_2 |}{\lambda}  \right)^{(t-\tau)}   \right)
\end{eqnarray}
where  $\boldsymbol{\rho}^r = [\rho_1^r, \ldots, \rho_n^r     ]^T$  is the right dominant eigenvector of the matrix $\mathbf{Q}$. This vector is normalized, $\sum_i \rho_i^l \rho_i^r =1$, and therefore, 
\begin{equation} \label{eq-product-dis}
\boldsymbol{\rho}_{prod} = \left[ \rho_1^l \rho_1^r, \ldots, \rho_n^l \rho_n^r     \right]^T 
\end{equation}
represents a quasi-stationary distribution of the absorbing Markov chain, which will be referred to as \textit{product distribution}. The left hand side of Eq.~(\ref{eq-q-distr-2}) converges to $\rho_j^l \rho_j^r$ as $\tau \to \infty$ and $t - \tau \to \infty$. 
Note that  $\{ \rho_i^l \rho_i^r  \}$ may be described as the distribution of the random variable at time $\tau$ ($\tau$ large), given that absorption has not yet taken place and will not take place for a long time.  The product distribution is more relevant than $\{ \rho_i^l \}$ in the sense that the Eq,~(\ref{eq-q-distr-1}) is a ``degenerate'' case of the Eq.~(\ref{eq-q-distr-2}).

Further, we look attentively at the structure of the global production network from the ``final use perspective'' and ``final value added perspective''. Here, we provide some basic definitions only for the final demand. A more detailed discussion for both perspectives will be presented in the Section~\ref{sect-results}. 
Let $m_{i \hat{j}  }$, where $i=1, \ldots, n$ and $\hat{j} = 1, \ldots, J$ be the probability that production from country-industry pair $i$ ends up as output purchased by the final users of country $\hat{j}$. We arrange the elements $m_{i \hat{j}  }$ as $n \times J$ matrix $\mathbf{M}$, which is called a \textit{global final demand matrix}. 
Let $\eta_{i \hat{j}} = \frac{f_{i, \hat{j}}}{x_i} $ and define $n \times J$ matrix $\mathbf{D}_\eta$ with elements $\eta_{i \hat{j}}$. In this case, Markov absorbing chain has $J$ absorbing states.  The transition matrix, Eq.~(\ref{eq-matrix-Pout}), now reads 
\begin{eqnarray} \label{eq-matrix-Pout-1}
{\mathbf{P}_{out}} &=& 
\begin{bmatrix}
\mathbf{I}_{J \times J} & \mathbf{0}_{J \times n} \\
\mathbf{D}_{\eta} & \mathbf{B} 
\end{bmatrix},
\end{eqnarray}
where $\mathbf{I}_{J \times J}$ is identity matrix and  $\mathbf{0}_{J \times n}$ zero matrix. A simple example for  illustrating all quantities proposed in this subsection is provided in Fig.~\ref{fig:example-1}. 

\section{Results and Discussions} \label{sect-results}

\begin{figure}[t!]
\centering
\includegraphics[width=8.7cm]{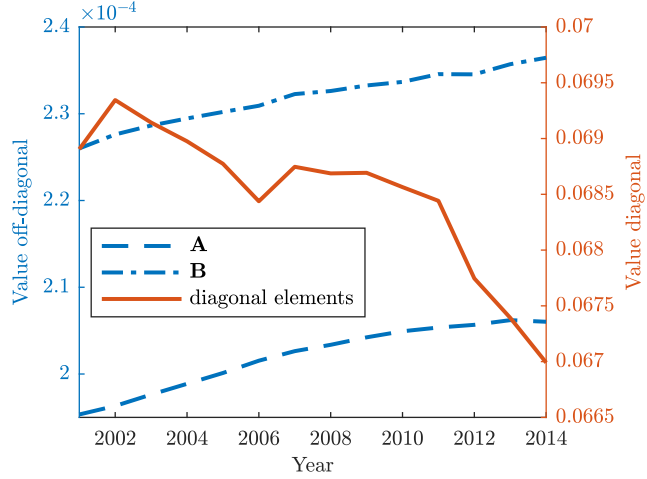}
\caption{Mean values for the off-diagonal blocks of $\mathbf{A}$ and  $\mathbf{B}$ (in blue). Mean values for the diagonal blocks of $\mathbf{A}$ and $\mathbf{B}$ (in orange).}
\label{fig:mean-value-off}
\end{figure}

The organization of world production is characterized by the structure of the input/output networks. 
%
%
%
Let us rewrite the $\mathbf{Z}$ matrix, Eq.~(\ref{eq-wiot-matrix-Z}), as follows:   
\begin{align} \label{eq-wiot-matrix-Z-1}
\mathbf{Z} &= \begin{bmatrix}
\mathbf{Z}_{11} & \ldots & \mathbf{Z}_{1J} \\
\vdots & \ddots & \vdots \\
\mathbf{Z}_{J1} & \ldots & \mathbf{Z}_{JJ} \\
\end{bmatrix}.
\end{align}
where each $\mathbf{Z}_{ij}$ is $S \times S$ matrix. Diagonal block matrices in Eq.~(\ref{eq-wiot-matrix-Z-1}) represent domestic inter-industry transactions, while off-diagonal blocks show the inter-country flows of intermediates via exports and imports. 
The adjacency matrices $\mathbf{A}$ and $\mathbf{B}$ are also rearranged in this way and their mean values are shown in Fig.~\ref{fig:mean-value-off}. The latter shows a growing fragmentation of production across countries worldwide.
From 2000 onwards, the global production networks became more complex and increasingly interconnected as a result of massively increased trade in intermediates (goods or services) among countries, both as buyers of foreign intermediate inputs (off-diagonal blocks of matrix $\mathbf{B}$) or sellers of intermediate products to third countries for further processing and export (off-diagonal blocks of matrix $\mathbf{A}$). 
It is noteworthy, however, that international fragmentation of production seems to have lost momentum in recent years, at least compared to 2000s, especially when it comes to forward GVC participation (many companies are probably less agile in responding to changing consumer demands). The consolidation of some value chains has been observed even during the 2008-2009 financial crisis, with some country-industry pairs switching back to domestic suppliers (slight increase in the mean values of the diagonal block in Fig.~\ref{fig:mean-value-off}) in the context of difficult access to trade finance and risks connected with international suppliers. Nevertheless, the recent slowdown may also point to a number of potential structural shifts facing the world economy (with many companies deciding to reexamine their outsourcing and production strategies), which could dramatically change the configuration of global production landscape and determine the future of globalization in a systemic way.
%



In input-output analysis, as mentioned before, upstreamness ($u_i$) and downstreamness ($d_i$) are used as measures of the importance of a certain country-industry pair $i$. 
In essence, larger $u_i$ values are related to relatively higher level of upstreamness of the particular country-industry pair $i$. The latter thus provides little to final consumers worldwide and instead sells disproportionately large share of its output (as intermediate inputs) to other producing industries in the world economy. Conversely, larger $d_i$ values are related to relatively higher level of downstreamness of a certain country-industry pair $i$. Clearly, the production process here relies disproportionately on intermediate inputs relative to the value-added from primary factors of production, and especially if purchases are made from those country-industry pairs which themselves use intermediate inputs intensively. 
That is, other things being equal, country-industry pairs with large $u_i$ or $d_i$ values are being  more proper targets for economic stimulation because they will bring much greater benefits to the entire world economy (by extending more of its resources to other country-industry pars in the former case, or by triggering other country-industry pairs to increase their outputs in the latter).
It should also be noted, however, that a country-industry with higher levels of downstreamness exhibit greater productivity fluctuations, because ``upstream supply-side shocks accumulate while propagating downstream'' \cite{mcnerney2018production}. Given the importance for policy makers, the next theorem provides mathematical framework for ordering (ranking) the country-industry pairs according to values of these measures (see Appendix A for the proof of the theorem).

\begin{figure}[t!]
\centering
\includegraphics[width=8.7cm]{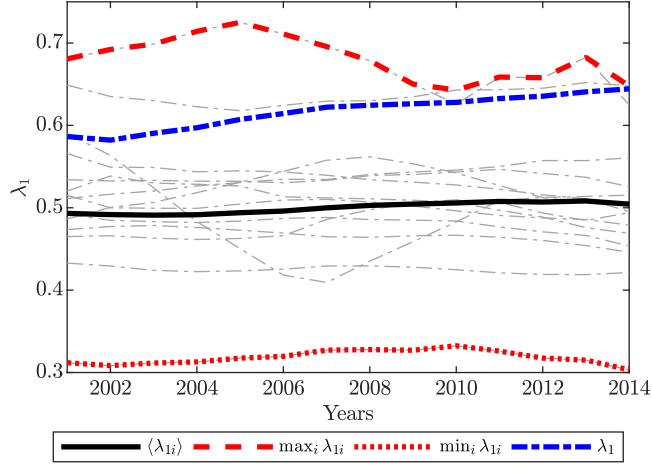}
\caption{The dominant eigenvalue of the world input/output network versus time. }
\label{fig:eigenvalues}
\end{figure}

\begin{theorem1} The following approximations hold:  
\begin{eqnarray}
\mathbf{u} &\approx & 
\left\{
\begin{array}{lll}
\mathbf{1}  +  \mathbf{B} \mathbf{1}   & \mbox{ for} &  \lambda \to 0^+  \\
 \frac{ \sum_i \rho_i^l(\mathbf{B})  }{1- \lambda}  \boldsymbol{\rho}^r(\mathbf{B}) & \mbox{ for} & \lambda \to 1^-
\end{array}
\right. \label{eq-app-u}    \\
\mathbf{d} &\approx & 
\left\{
\begin{array}{lll}
\mathbf{1}  +   \mathbf{A}^{T} \mathbf{1}  & \mbox{ for} & \lambda \to 0^+ \\
\frac{ \sum_i \rho_i^r(\mathbf{A}) }{1- \lambda} \boldsymbol{\rho}^l(\mathbf{A}) & \mbox{ for} & \lambda \to 1^-  
\end{array}
\right.  \label{eq-app-d} 
\end{eqnarray}
where $\boldsymbol{\rho}^l(\mathbb{\cdot})$  and  $\boldsymbol{\rho}^r(\cdot)$ are  the left and the right dominant eigenvectors of the matrices $\mathbf{A}$ and $\mathbf{B}$. 
%
%
\end{theorem1}

\begin{figure}[t!]
\centering
\includegraphics[width=8.7cm]{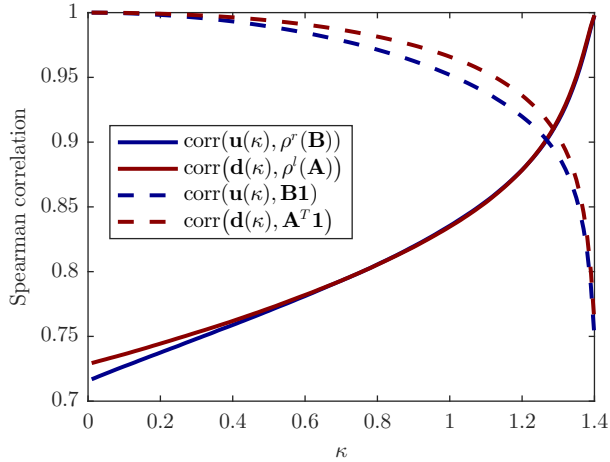}
\caption{Spearman’s rank correlation for the WIOD, year 2014. Upstreamness     $\mathbf{u}(\kappa)$ and downstreamness $\mathbf{d}(\kappa)$   are computed as
$ \mathbf{u}(\kappa ) = ( \mathbf{I} - \kappa \mathbf{B} )^{-1} \mathbf{1}$ and  $ \mathbf{d}(\kappa ) = ( \mathbf{I} - \kappa \mathbf{A}^T )^{-1} \mathbf{1}$, respectively, where $\kappa$ is a parameter. 
$\kappa=1$ corresponds to the values of the $\mathbf{u}$ and $\mathbf{d}$ of the world networks and the year 2014.  The limits $\kappa \to 0^+$ and $\kappa \to (1/\lambda)^-$ correspond to $\lambda \to 0^+$ and $\lambda \to 1^-$, respectively (see Appendix A).}
\label{fig:correlation}
\end{figure}

Note that in Eq.~(\ref{eq-app-u})  and Eq.~(\ref{eq-app-d}),   $\mathbf{B} \mathbf{1}$ and $ \mathbf{A}^T \mathbf{1}$ are vectors of out-degree and in-degree of the nodes of the output-network and input-network, respectively.    
Therefore, when $\lambda \to 0$, it follows that the ordering (or the ranking) of the country-industry pairs depends only on the out-degree (in-degree) centrality, and thus, high-degree nodes are important, that is, the most important suppliers of value added in world GVCs  
(or the general-purpose country-industry pairs whose value added contained in intermediate inputs is sent to a wide range of country-industry pairs for further processing) and the country-industry pairs that import intermediates from many sources (or the most important recipients of foreign value added in global production networks).
On the other hand, when $\lambda \to 1$, the ordering (ranking) depends only on the right (left) dominant eigenvector of the matrix $\mathbf{B}$ ($\mathbf{A}$).
Therefore, in this case, it could happen that a low-degree country-industry pair has greater influence (i.e. it is more relevant player in the global production networks) than high-degree hub. Put differently, the centrality of a country-industry pair here is recurrently related to the pairs to which it is connected, that is a node’s position depends on the importance of its neighbors (a node eigenvector centrality).  

Figure~\ref{fig:eigenvalues} shows the dominant eigenvalue of the world input/output network as well as the country dominant eigenvalues of the input/output networks of all countries versus time. The mean value of dominant eigenvalues of the input/output networks of all countries, $ \langle \lambda_{1i} \rangle $, is almost constant in time, while the  dominant eigenvalue of the world input/output network, $\lambda_1$, increases in time. Moreover, figure~\ref{fig:correlation} depicts Spearman’s rank correlation between upstreamness (downstreamness) and  the out-degree (in-degree) centrality and  right dominant eigenvector of the adjacency matrix $\mathbf{B}$ ($\mathbf{A}$) for the WIOD. In the limits $\kappa \to 0^+$ and $\kappa \to (1/\lambda)^-$ (that correspond to the limits $\lambda \to 0^+$ and $\lambda \to 1^-$, respectively, see the Appendix A), the orderings of $\mathbf{u}$ and $\mathbf{d}$ are perfectly matched with the ordering of the vectors from the right-hand side of the Eq.~(\ref{eq-app-u}) and Eq.~(\ref{eq-app-d}), respectively. When $\kappa=1$, the ordering of $\mathbf{u}$ and $\mathbf{d}$ is 95\% correlated with the ordering of degree centrality and 80\% with eigenvectors' ordering.

\begin{figure}[t!]
\centering
\includegraphics[width=8.7cm]{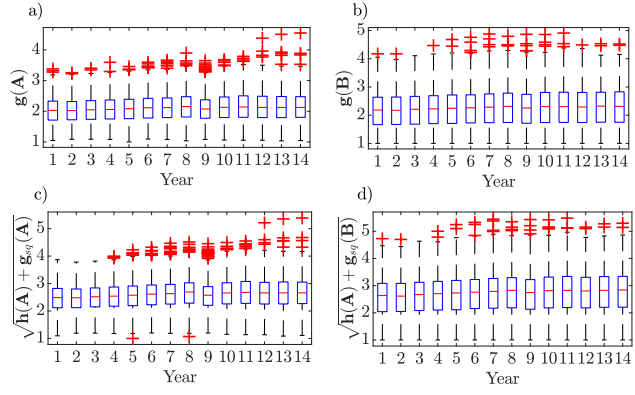}
\caption{Comparative boxplots of $\mathbf{g}$ and $\mathbf{h}$. }
\label{fig:boxplot}
\end{figure}

Two random variables have been proposed in the previous section. Assume that the process starts in state $i$.  As already stated, the first random variable, $X_{ij}$, represents the number of visits to state $j$ before absorption, while  the second,  $X_i$ is  the time to  absorption. Expectation and variance of these random variables are provided by Eq.~(\ref{eq-mean-Q}),  Eq.~(\ref{eq-var-Q}), Eq.~(\ref{eq-mean-X}), and Eq.~(\ref{eq-var-X}), respectively. 
Figure~\ref{fig:boxplot} provides a visual display of summarizing a distribution of data, or comparative boxplots that can be used to compare the distributions of expectation and variance of the time to absorption for both the world-input and world-output networks. The most noticeable feature is that expectation, or expected number of steps before absorption, and variance of the number of steps before absorption for the output network (and the respective upstreamness) are generally higher than those for the input network (and the respective downstreamness). Put differently, the median and the upper quartile for the former sample are all above the corresponding values for the latter sample (the lower quartiles are roughly similar). The dispersion is also greater for the output network, that is the interquartile range, as revealed by the box lengths, is reasonably longer and so is the overall range of dataset (with or without `outliers').

The next result unveils that both the input chain and the output chain have exactly the same quasi-stationary product distribution (for the proof of the theorem, see Appendix B). 


\begin{theorem1} The World input/output network is characterized with 
\begin{equation}
\rho_i^l(\mathbf{A}) \rho_i^r(\mathbf{A})  = 
\rho_i^l(\mathbf{B})  \rho_i^r(\mathbf{B})  \mbox{ for all } i \in V. 
\end{equation}
\end{theorem1}

Note that although $\rho_i^l(\mathbf{A}) \neq  \rho_i^l(\mathbf{B}) $  and $  \rho_i^r(\mathbf{A}) \neq \rho_i^r(\mathbf{B})$, their products are equal to each other. Therefore, the product distribution does not depend on the type of network (input/output). This, rather surprising, fact runs counter to what might be expected at a time of rising importance for the global value chains.

\begin{figure}[t!]
\centering
\includegraphics[width=8.7cm]{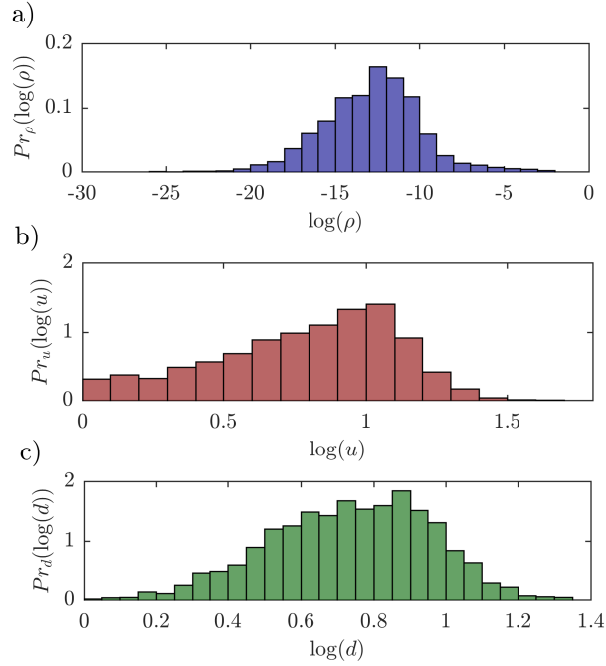}
\caption{ Histograms (WIOD, year 2014) of (a) the product distribution, (b) the output upstreamness, and (c) the input downstreamness.  The product distribution does not depend on the type of network (input/output). }
\label{fig:histogram}
\end{figure}

In order to explain this surprising (unexpected) result, we interpret the probability  $ \rho_j^l \rho_j^r$ as limiting conditional mean ratios. It can be shown that: 
\begin{eqnarray}
\sum_{i \in V} \pi_i \mathbb{E} \left[  \left. \frac{X_{ij}}{X_i} \right| X_i = t      \right]  & = & \rho_j^l \rho_j^r + O \left( \frac{1}{t} \right) \label{eq-ratio-1} \\ 
\sum_{i \in V} \pi_i \mathbb{E} \left[  \left. \frac{ \sum_{k=0}^t Y_{ij}^{(k)} }{t} \right| X_i > t      \right]  & = & \rho_j^l \rho_j^r + O \left( \frac{1}{t} \right) \label{eq-ratio-2}
\end{eqnarray}
The left hand side of Eq.~(\ref{eq-ratio-1}) is the expected proportion of time spent in the state $j$ before absorption. 
Since the transaction flows of intermediates coming from (or sold by) the pair $j$ (world input network) are, at the same time, equal to flows of intermediates purchased by the pair $j$ (world output network), the expected proportion of time spent in the state $j$ before absorption for the input network is equal to  the expected proportion of time spent in the state $j$ before absorption for the output network. 
Moreover, note that $ \sum_{k=0}^t Y_{ij}^{(k)}$, see Eq.~(\ref{eq-ratio-2}),  equals the number of visits to state $j$ up to time $t$. Therefore, the expected proportion of time spent in the state $j$ before absorption is equal to $\rho_j^l \rho_j^r$ and does not depend on the type of the network (world input/world output network).
Figure~\ref{fig:histogram} depicts the histograms of the product distribution $ \rho_j^l \rho_j^r$, the output upstreamness $\mathbf{u}$,  and the input downstreamness $\mathbf{d}$ for the WIOD and the year 2014. 

The next result is actually a combination of the classical results in respect of the absorbing Markov chains \cite{kemenyfinite} and some new insights (for the proof, see Appendix C).  
\begin{theorem1}
(i) The global final demand matrix  $\mathbf{M}$   can be computed as 
%
\begin{eqnarray}
\mathbf{M} &=& ( \mathbf{I} - \mathbf{B}  )^{-1} \mathbf{D}_\eta    \label{eq-matrix-M}  
\end{eqnarray}
such that  $\mathbf{M}$ is a row stochastic matrix.  \\
\noindent
(ii)
\begin{eqnarray}
( \mathbf{I} - \mathbf{A}^T  )^{-1} \boldsymbol{\delta} & = & \mathbf{1} \label{eq-distr-value-added}
\end{eqnarray} 
\\
\noindent
(iii) For an economy for which  the row sums of $\mathbf{A}^T$  are all equal to $c$,  the input downstreamness is constant $\mathbf{d} = \frac{1}{1-c} \mathbf{1}$. If the row sums of $\mathbf{B}$  are all equal to $c$, then  the
the output upstreamness is constant $\mathbf{u} = \frac{1}{1-c} \mathbf{1}$.  
%
\end{theorem1}

\begin{figure}[t!]
\centering
\includegraphics[width=8.7cm]{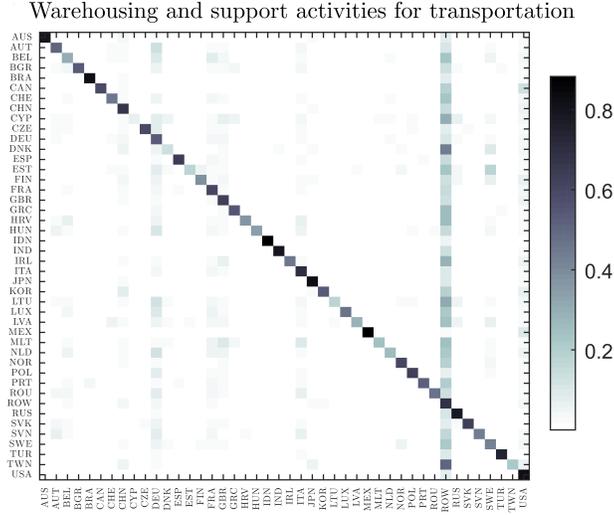}
\caption{Global patterns of final demand for  the  warehousing and support activities for transportation in 2014.}
\label{fig:55}
\end{figure}

Note that the matrix $\mathbf{D}_\eta$ is the one-step transition matrix, while the vector (for a fixed $i$) 
\begin{eqnarray}
\mathbf{m}_i & = & [m_{i1}, \ldots, m_{iJ} ]^T  \label{eq-vector-m}  
\end{eqnarray}
is a probability distribution showing how the final output of country-industry pair $i$ is distributed among different countries.
For a fixed industry $r$ we arrange the elements $m_{(\hat{i},r),\hat{j} } $ in a $J\times J$ matrix, 
\begin{equation} \label{eq-matrix-PP}
\mathbf{PP}= \left[ m_{(\hat{i},r),\hat{j} } \right], 
\end{equation}
which we call a \textit{global final demand matrix  of the industry} $r$, showing the global patterns of final demand for any given industry.
Similarly, if we  write $\mathbf{L} = [ \ell_{ij}] \equiv ( \mathbf{I} - \mathbf{A}^T  )^{-1} $ and define  $\zeta_{i \hat{j}} = \sum_{r=1}^S  \ell_{i (\hat{j} r)} \delta_{ \hat{j} r }$, it follows from Eq.~\ref{eq-distr-value-added} that the vector (for a fixed $i$) 
\begin{eqnarray}
 \boldsymbol{\zeta}_i & = &  [ \zeta_{i1}, \ldots, \zeta_{iJ} ]^T \label{eq-zeta}
\end{eqnarray}
is the  probability distribution showing how the value added of country-industry pair $i$ is distributed among different countries.
Again,   for a fixed industry $r$ we arrange the elements $\zeta_{(\hat{i},r),\hat{j} } $ in a $J\times J$ matrix, 
\begin{equation} \label{eq-matrix-WP}
    \mathbf{WP}= \left[ \zeta_{(\hat{i},r),\hat{j} }  \right], 
\end{equation}
which we call a \textit{ global value added matrix of the industry $r$}, capturing the final impact after all stages of production have circulated throughout the world economy and showing the global patterns of value added for any given industry.
This metric, which breaks down the distribution of gross trade flows along the sources and destinations of value added, provides a coherent answer to many important questions about the interconnections among countries, especially with regard to the aggregate impact and propagation of shocks. For example, the important role that particular countries play in international flows of value added raises the questions about the resilience of the world trading system if they suffer a large-scale economic shock. All of this has a major impact on forecasting the macroeconomic developments and on monetary policy decisions.     
Figure~\ref{fig:55} shows global  demand matrix for the warehousing and support activities for transportation in 2014. Each row (a horizontal line) is a probability distribution: the $j$ element of the row $i$ shows the probability that a good/service produced in the country $i$ has been delivered to final consumers in country $j$. 
%
On the other, the column $j$ (a vertical line) represents the final consumer buying patterns for a particular product (e.g. warehousing and support activities for transportation) in country $j$.

\begin{figure}[t!]
\centering
\includegraphics[width=8.7cm]{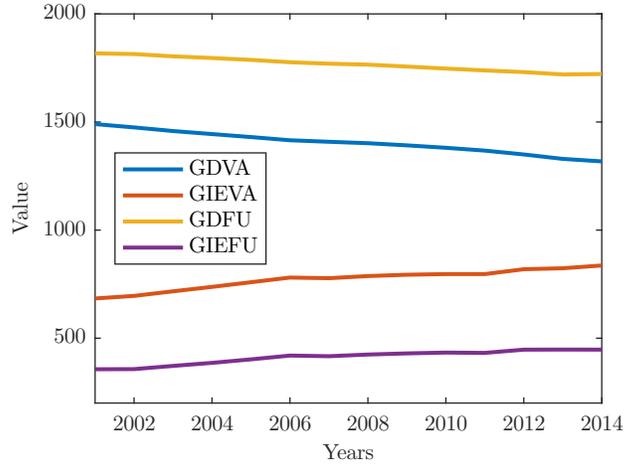}
\caption{Global quantities over years.}
\label{fig:global-quantities}
\end{figure}

Finally, the availability of global input-output matrices has paved the way to several methodological contributions on measures of trade in value added, or proxies of GVC participation (see \cite{amador2017networks} for a review). In view of diagonal and off-diagonal blocks of our Global Value-Added and Global Final Demand Matrices, the proposed metrics below provide some new insights, or novel proxies for the analysis of global production chains:  
\begin{eqnarray}
\mbox{GDVA} & = &  \sum_{\hat{i} =1}^J \sum_{r=1}^S \zeta_{\hat{i},r ; \hat{i}  } \label{eq-gdva} \\
\mbox{GIEVA} & = &\sum_{\hat{i} =1}^J \sum_{\hat{j} =1, \hat{j} \neq \hat{i}  }^J   \sum_{r=1}^S \zeta_{\hat{i},r ; \hat{j}  } \label{eq-gieva}  \\
\mbox{GDFU} & = &  \sum_{\hat{i} =1}^J \sum_{r=1}^S m_{\hat{i},r ; \hat{i}  } \label{eq-gdfu} \\
\mbox{GIEFU} & = &\sum_{\hat{i} =1}^J \sum_{\hat{j} =1, \hat{j} \neq \hat{i}  }^J   \sum_{r=1}^S m_{\hat{i},r ; \hat{j}  } \label{eq-giefu}
\end{eqnarray}
The first indicator, which we call Global Domestic Value-Added (GDVA), refers to all domestic value-added flows within a given country-industry pair, irrespective of the number of steps along the chain (see Eq. (\ref{eq-gdva})). Hence, the measure includes both, direct and indirect domestic value-added contents. The former denotes the income of primary inputs directly involved in production (one stage), while the latter encompasses both the domestic flows of intermediate goods and services across industries (hence, the magnitude only depends on the density of domestic intersectoral linkages) and domestic value-added that is re-imported in the economy of origin as a part of other intermediates (hence, the magnitude of value added depends on the density of intersectoral linkages between two or more countries).       
The second metric we call Global Import-Export Value-Added (GIEVA) indicates the extent to which a country-industry pair is connected to global production networks for its foreign trade (see Eq. (\ref{eq-gieva})). Hence, this measure refers to inter-country transaction flows of intermediate products (goods and services) through exports and imports.   
The third indicator, or Global Domestic Final Use (GDFU), indicates the final use of domestic output by the country itself (see Eq. (\ref{eq-gdfu})).
The last indicator, which we call Global Import-Export Final Use (GIEFU), refers to all imports and exports destined for final consumers worldwide (see Eq.~(\ref{eq-giefu})). 
The evidence suggests that global production networks have provided a great impulse to globalization during the past decades (Figure \ref{fig:global-quantities}). After the marked growth during the early 2000s, international fragmentation of production has become a cornerstone of the global economy, with products effectively being ``made in the world''. Growing fragmentation, with a short disruption during the 2008 financial crisis, is typically associated with more trade in intermediate inputs via exports and imports (an overall upward trend in GIEVA) and less domestic value added (noticeable decline in GDVA) (Figure \ref{fig:global-quantities}). Nevertheless, global production networks, at least compared to early 2000s, seems to have lost momentum during the last years (Figure \ref{fig:global-quantities}). In point of fact, the world production has been made more expensive since the economic crisis, mainly due to increasing trade costs and protectionism. Moreover, the recent steady state in GDFU and GIEFU (Figure \ref{fig:global-quantities}) points to a somewhat greater attractiveness of localized production and potentials for shortening the production chain lengths which means that individual parts and products would increasingly be manufactured in proximity to final users. It is clear that such dynamics in international fragmentation will promptly have impact on the world economy and probably give rise to a certain re-configuration of the global production networks.

\section{Conclusions} 

This paper aims to contribute to our understanding of global production networks and international transmission of economic shocks. Here, by virtue of recently available world input-output database, we propose discrete-time absorbing Markov chains to model the structure and interdependence among country-industry pairs of the world economy. The examination of various properties of the constructed Markov chains allows for a more in-depth analysis of the underlying dynamical system than any currently available. In this context, we have introduced three random variables and set them against the existing approaches (upstreamness and downstreamness) to measure the positioning of countries and industries in GVCs: (1) The time to absorption from the seller perspective (output supply chain) or supply side in global production networks, that is the number of times a particular country-industry pair contributes (domestic value added contained in intermediates sent to a partner economy for further processing and export) along the route of each unit of value added in the global economy before reaching the final consumers; (2) The time to absorption from the buyer perspective (input demand chain) or sourcing side in global production networks, that is the expected number of steps it takes primary inputs to reach a certain country-industry pair, or number of times the intermediate country-industry pair(s) contribute(s) in total input of a particular country-industry pair that incorporates foreign value added (or imports intermediate inputs to produce goods/services that are subsequently exported for final consumption or intermediate use); (3) The time spent in a state before absorption or the number of times a certain country-industry pair becomes an intermediate in its own supply or demand chains before absorption. Based on these variables, we have proposed several quantities that are summarized in Table~\ref{table:1}. More specifically, our measure of the variance of time to absorption can help the policy-makers to develop better models of production. Besides, we have shown that both the input and output chains exhibit exactly the same quasi-stationary product distribution. Next, our probability distributions and the related global final demand/global value-added matrices provide valuable information about the global patterns of final demand or the trade in value-added and interconnections among countries, especially with regard to the propagation of economic shocks. Finally, on the basis of these matrices, we have proposed several novel proxies for participation and analysis of world production chains. The application of these measures can help assess the structural shifts in the world economy that may possibly change the configuration of global production landscape and  shape the future of globalization.

\begin{table}[t!]
\centering
\begin{tabular}{ | m{1cm} | m{3cm}| m{2.5cm} | m{2cm} | } 
\hline
 & quantity & computed with   & equation  \\ 
\hline
$\mathbf{g}(\mathbf{B})$  & $n$-dim vector  &  $\mathbf{B}$ &  (\ref{eq-mean-X})  \\ 
\hline
$\mathbf{g}(\mathbf{A})$ & $n$-dim vector & $\mathbf{A}$  & (\ref{eq-mean-X}) \\ 
\hline
$\mathbf{h}(\mathbf{B})$  & $n$-dim vector  &  $\mathbf{B}$ &  (\ref{eq-var-X})  \\ 
\hline
$\mathbf{h}(\mathbf{A})$ & $n$-dim vector & $\mathbf{A}$  & (\ref{eq-var-X}) \\ 
\hline
$\boldsymbol{\rho}_{prod}$ & $n$-dim vector & $\mathbf{A}$ or $\mathbf{B}$ & (\ref{eq-product-dis}) \\ 
\hline
$\mathbf{m}_i$  & $J$-dim vector & $\mathbf{B}$ and $\mathbf{D}_\eta $ &  (\ref{eq-vector-m})   \\ 
\hline
$ \boldsymbol{\zeta}_i $ & $J$-dim vector & $\mathbf{A}$ and $\boldsymbol{\delta}$ & (\ref{eq-zeta}) \\ 
\hline
$ \mathbf{PP} $ & $J \times J$ matrix & $\mathbf{B}$ and $\mathbf{D}_\eta $  & (\ref{eq-matrix-PP}) \\ 
\hline
$ \mathbf{WP} $ & $J \times J$ matrix & $\mathbf{A}$ and $\boldsymbol{\delta}$ & (\ref{eq-matrix-WP}) \\ 
\hline
GDVA  & scalar  &  $\mathbf{A}$ and $\boldsymbol{\delta}$ &  (\ref{eq-gdva})  \\ 
\hline
GIEVA & scalar & $\mathbf{A}$ and $\boldsymbol{\delta}$  & (\ref{eq-gieva}) \\ 
\hline
GDFU  & scalar  &  $\mathbf{B}$ and $\mathbf{D}_\eta $   &  (\ref{eq-gdfu})  \\ 
\hline
GIEFU & scalar & $\mathbf{B}$ and $\mathbf{D}_\eta $  & (\ref{eq-giefu}) \\ 
\hline
\end{tabular}
\vspace{0.5cm}
\caption{Novel quantities proposed in the manuscript}
\label{table:1}
\end{table}

\section*{Appendix A: Proof of the Theorem 1} 

Let  $ \mathbf{u}(\kappa ) = ( \mathbf{I} - \kappa \mathbf{B} )^{-1} \mathbf{1}$ and  $ \mathbf{d}(\kappa ) = ( \mathbf{I} - \kappa \mathbf{A}^T )^{-1} \mathbf{1}$, where $\kappa > 0 $ is a parameter such that $\kappa < 1/\lambda $.  
The first main result states that the vectors $\mathbf{u}(\kappa)$ and $\mathbf{d}(\kappa)$ can be approximated as  
\begin{eqnarray}
\mathbf{u}(\kappa) &\approx & 
\left\{
\begin{array}{lll} 
\mathbf{1}  +  \mathbf{B} \mathbf{1},   & \mbox{ for} &  \kappa \to 0^+,  \\
 \frac{ \boldsymbol{\rho}^l(\mathbf{B})^T \mathbf{1}  }{1- \kappa \lambda}  \boldsymbol{\rho}^r(\mathbf{B}), & \mbox{ for} & \kappa \to \frac{1}{\lambda}^-,
\end{array}
\right. \label{eq-app-u-1}    \\
\mathbf{d}(\kappa) &\approx & 
\left\{
\begin{array}{lll}
\mathbf{1}  +   \mathbf{A}^T \mathbf{1},  & \mbox{ for} & \kappa \to 0^+, \\
\frac{\mathbf{1}^T \boldsymbol{\rho}^r(\mathbf{A})}{1-\kappa \lambda}  \boldsymbol{\rho}^l(\mathbf{A}) & \mbox{ for}, & \kappa \to \frac{1}{\lambda}^-, 
\end{array}
\right.  \label{eq-app-d-1} 
\end{eqnarray}
where $\boldsymbol{\rho}^l(\mathbf{M})$ and $\boldsymbol{\rho}^r(\mathbf{M})$ are the left and right eigenvectors associated with the largest eigenvalue of $\mathbf{M}$, 
respectively. In what follows, we first provide a detailed background on relevant properties of positive matrices and functions of matrices. Afterwards, we give a full proof of the theorem. 


\subsection*{Perron-Frobenius theorem for positive matrices}
 
Recall some known properties of positive matrices. Let $\mathbf{M} = [M_{{ij}}]$ be an $N\times N$ positive matrix: $a_{{ij}}>0$ for $1\leq i,j\leq N$. Then the following statements hold.
\begin{itemize}
\item 

There is a positive real number $\lambda_1$, (called the Perron root, the Perron-Frobenius eigenvalue, the leading eigenvalue,  or the dominant eigenvalue), such that $\lambda_1$ is an eigenvalue of $\mathbf{M}$ and any other eigenvalue (possibly, complex) in absolute value is strictly smaller than $\lambda_1$, 
\begin{equation} \label{eq-lambdai-lambda1}
\mid \lambda_i \mid < \lambda_1, 
\end{equation}
for $i=2, \ldots, N$. 

\item 
$\lambda_1$  is a simple root of the characteristic polynomial of $\mathbf{M}$. 

\item 
There exists a right eigenvector $\boldsymbol{\rho}^r  = [\rho^r_1, \ldots, \rho^r_N]^T$ of $\mathbf{M}$ with eigenvalue $\lambda_1$ such that 
$\mathbf{M}\boldsymbol{\rho}^r = \lambda_1 \boldsymbol{\rho}^r$, $\rho^r_i > 0$ for $i=1, \ldots, N$. 
Respectively, there exists a positive left eigenvector $\boldsymbol{\rho}^l = \left[ \rho^l_1, \ldots, \rho^l_N  \right]^T$ such that $\left[ \boldsymbol{\rho}^l \right]^T \mathbf{M} = \lambda_1 \left[ \boldsymbol{\rho}^l \right]^T$ and $\rho^l_i > 0$ for all $i$. 

\end{itemize}

%
%

\subsection*{Spectral Theorem for Diagonalizable Matrices} 

An $N \times N$ matrix $\mathbf{M}$ with spectrum $\sigma(\mathbf{M}) = \{\lambda_1, \lambda_2, \ldots, \lambda_N \} $ is said to diagonalizable if and only if there exist matrices $\{ \mathbf{G}_1, \mathbf{G}_2, \ldots , \mathbf{G}_N \}$ such that 
\begin{align} \label{eq-spec-dec}
\mathbf{M}= \lambda_1 \mathbf{G}_1 + \lambda_2 \mathbf{G}_2 + \ldots + \lambda_N \mathbf{G}_N, 
\end{align}
where the $\mathbf{G}_i$'s have the following properties: (i) $\mathbf{G}_i$ is the projector onto $K (\mathbf{M} - \lambda_i \mathbf{I})$ along $R( \mathbf{M} - \lambda_i \mathbf{I})$; (ii) $\mathbf{G}_i \mathbf{G}_j = 0  $ whenever $i \neq j$; and (iii) $ \mathbf{G}_1 + \mathbf{G}_2  + \ldots + \mathbf{G}_N = \mathbf{I}$.  The expansion~(\ref{eq-spec-dec}) is known as the spectral decomposition of $\mathbf{M}$, and the $\mathbf{G}_i$'s are called the spectral projectors associated with $\mathbf{M}$.
Moreover, if $\boldsymbol{\rho}^r$ and $\boldsymbol{\rho}^l$ are the respective right-hand and left-hand eigenvectors associated with a simple eigenvalue $\lambda$, then
spectral projector associated with $\lambda$ is: 
\begin{align} \label{eq-right-left}
\mathbf{G} = \frac{\boldsymbol{\rho}^r \left[ \boldsymbol{\rho}^l \right]^T}{\left[ \boldsymbol{\rho}^l \right]^T \boldsymbol{\rho}^r}.
\end{align}
%
If $\mathbf{M}$ is a diagonalizable matrix, then it is also similar to a diagonal matrix $\mathbf{D}$. Note that two $N \times N$ matrices $\mathbf{M}$ and  $\mathbf{D}$ are said to be similar whenever there exists a nonsingular matrix $\mathbf{P}$ such that $\mathbf{P}^{-1} \mathbf{M} \mathbf{P} = \mathbf{D}$.

\subsection*{Functions of Matrices}

Let $\mathbf{M} = \mathbf{P} \mathbf{D} \mathbf{P}^{-1}$ be a diagonalizable matrix where the eigenvalues in $\mathbf{D}  = \mbox{diag} \left( \lambda_1 \mathbf{I}, \lambda_2 \mathbf{I}, \ldots, \lambda_N \mathbf{I} \right)$ are grouped by repetition. For a function $f(z)$ that is defined at each $\lambda_i$, define
\begin{eqnarray}
f(\mathbf{M}) & = &  \mathbf{P} f(\mathbf{D})  \mathbf{P}^{-1} \nonumber \\
& = &  \mathbf{P}  
\begin{bmatrix}
    f(\lambda_1) \mathbf{I}     & 0 & 0 & \dots & 0 \\
    0      & f(\lambda_2) \mathbf{I} & 0 & \dots & 0 \\
    \hdotsfor{5} \\
    0      & 0 & 0 & \dots & f(\lambda_N) \mathbf{I}
\end{bmatrix}
\mathbf{P}^{-1}  \nonumber \\
& = & f(\lambda_1) \mathbf{G}_1 + f(\lambda_2) \mathbf{G}_2  + \ldots + f(\lambda_N)\mathbf{G}_N. \label{eq-fA-for-d}
\end{eqnarray}

We now briefly discuss functions of nondiagonalizable matrices following Ref.~\cite{meyer2000matrix}.
For an arbitrary matrix $\mathbf{M} \in \mathbb{C}^{N \times N}$ with $\sigma(\mathbf{M}) = \{ \lambda_1, \ldots, \lambda_s\}$ where  $s$ is the number of distinct eigenvalues of $\mathbf{M}$, let  $k_i$ be the index of the eigenvalue $\lambda_i$ (that is, the order of the largest Jordan block associated with $\lambda_i$ in the Jordan canonical form of $\mathbf{M})$.
A function $f: \mathbb{C} \to \mathbb{C}$ is said to be defined (or to exist) at $\mathbf{M}$ when $f(\lambda_i), f^{\prime}(\lambda_i), \ldots, f^{(k_i-1)}(\lambda_i)$ exist for each $\lambda_i$. 
%
%
If $f$ exists at $\mathbf{M}$, then the value of $f$ at $\mathbf{M}$ is defined to be
\begin{eqnarray}
%
%
f(\mathbf{M}) &=& \sum_{i=1}^s \sum_{j=0}^{ k_i - 1 } \frac{f^{(j)}(\lambda_i) }{j!}  \left( \mathbf{M}  - \lambda_i \mathbf{I}  \right)^{j} \mathbf{G}_i. \label{eq-Spectral-Resolution}
\end{eqnarray}






For an arbitrary square matrix $\mathbf{M}$, as a particular example of $f$ we consider the geometric series $f(z) = 1+z+z^2 + \ldots $ also known as the Neumann series 
$$
f(\mathbf{M}) = \sum_{k=0}^\infty  \mathbf{M}^k. 
$$ 
Let $\rho(\mathbf{M})$ be the spectral radius of $\mathbf{M}$  defined as $ \rho(\mathbf{M}) = \max_{ \lambda \in \sigma(\mathbf{M})} \mid \lambda \mid$.
The following statements are equivalent: (i) The Neumann series converges. 
(ii) $ \rho(\mathbf{M}) < 1$, and (iii) $\lim_{k \to \infty}  \mathbf{M}^k = 0$. 
In which case, $(\mathbf{I} -  \mathbf{M})^{-1}$ exists and
$$
\sum_{k=0}^\infty  \mathbf{M}^k = (\mathbf{I} - \mathbf{M})^{-1}.
$$

\subsection*{Proof of the theorem}

Let us now consider an arbitrary positive matrix $\mathbf{M}$. Let $\lambda_1, \ldots, \lambda_N \in \mathbb{C}$ be its eigenvalues. From the Perron-Frobenius theorem it follows that $\lambda_1 > \mid \lambda_2 \mid \geq \ldots \geq \mid \lambda_N \mid$, we consider the parametrized vector $\mathbf{g}$  defined as 
\begin{equation} \label{eq-world-index-with-t}
\mathbf{g}(\kappa) =  f( \kappa \mathbf{M}) \mathbf{1} =  \left(  \mathbf{I} - \kappa \mathbf{M}      \right)^{-1} \mathbf{1}.
\end{equation}
where $\kappa > 0$ is a real parameter such that $\kappa < 1/\lambda_1$.  
Let $ \boldsymbol{\rho}^r(\mathbf{M})$ be the dominant right eigenvector of $\mathbf{M}$  and let $ \boldsymbol{\rho}^l(\mathbf{M})$ be the dominant left eigenvector of $\mathbf{M}$.
Assume that $\mathbf{M}$ is diagonalizable matrix. Then by combining  (\ref{eq-fA-for-d}) and (\ref{eq-right-left}) we have 
\begin{eqnarray*}
\mathbf{g}(\kappa) & = & \left[ f(\kappa \lambda_1) \mathbf{G}_1 + f(\kappa \lambda_2) \mathbf{G}_2  + \ldots + f(\kappa \lambda_N)\mathbf{G}_N  \right]  \mathbf{1} \\
& = & \left[ f(\kappa \lambda_1) \frac{\boldsymbol{\rho}^r \left[ \boldsymbol{\rho}^l \right]^T}{\left[ \boldsymbol{\rho}^l \right]^T \boldsymbol{\rho}^r} + f(\kappa \lambda_2) \mathbf{G}_2  + \ldots + f(\kappa \lambda_N)\mathbf{G}_N  \right] \mathbf{1}. 
\end{eqnarray*}
Assuming $\left[ \boldsymbol{\rho}^l \right]^T \boldsymbol{\rho}^r = 1$ (normalization) and since $f(\kappa \lambda_1) >0$, $\boldsymbol{\rho}^l > 0$ and $\mbox{\boldmath$\beta$ } > 0$ from the last equation it follows that
$$
\frac{\mathbf{g}(\kappa) }{ f(\kappa \lambda_1) \left[ \boldsymbol{\rho}^l \right]^T   \mbox{\boldmath$\beta$ }}
=    \boldsymbol{\rho}^r + \frac{f( \kappa \lambda_2)}{ f(\kappa \lambda_1) \left[ \boldsymbol{\rho}^l \right]^T \mathbf{1} }   \mathbf{G}_2 \mathbf{1}  + \ldots + \frac{f(\kappa \lambda_N)}{ f(\kappa \lambda_1) \left[ \boldsymbol{\rho}^l \right]^T  \mathbf{1}  }  \mathbf{G}_N  \mathbf{1}.
$$
Let us examine the two limiting cases as $\kappa \to \frac{1}{\lambda_1}$ and $\kappa \to 0$. 

First, since for $z=1$, the series $f(z)$ diverges, as $\kappa \to \frac{1}{\lambda_1}$, the denominator of the right-hand side of the last  equation approaches infinity. On the other hand, each derivative $f^{(j)}(z)$ of $f(z)$ can be expressed by a power series having the same radius of convergence as the power series expressing $f(z)$. From Eq.~(\ref{eq-lambdai-lambda1}), it follows that
 $ \bigl| f^{(j)}\left( \frac{\lambda_i}{\lambda_1} \right) \bigr|  < \infty$, and hence,
$$
\lim_{\kappa \to \frac{1}{\lambda_1}^{-}}  \frac{\mathbf{g} (\kappa) }{ f(\kappa \lambda_1) \left[ \boldsymbol{\rho}^l \right]^T  \mathbf{1}  }
=   \boldsymbol{\rho}^r.
$$
More generally, for nondiagonalizable matrices, from (\ref{eq-Spectral-Resolution}) it follows  
\begin{eqnarray}
\mathbf{g}(\kappa) &=& f( \kappa \lambda_1) \boldsymbol{\rho}^r \left[ \boldsymbol{\rho}^l \right]^T \mathbf{1}   + \sum_{i=2}^s \sum_{j=0}^{ k_i - 1 } \frac{f^{(j)}(\kappa \lambda_i) }{j!} \left( \mathbf{M}  - \lambda_i \mathbf{I}  \right)^{j} \mathbf{G}_i \mathbf{1}   \nonumber \\
\frac{ \mathbf{g}(\kappa)}{f(\kappa\lambda_1) \left[ \boldsymbol{\rho}^l \right]^T \mathbf{1} } &=&  \boldsymbol{\rho}^r + \sum_{i=2}^s \sum_{j=0}^{ k_i - 1 } \frac{1}{\left[ \boldsymbol{\rho}^l \right]^T \mathbf{1} } \frac{f^{(j)}(\kappa \lambda_i) }{j! f ( \kappa \lambda_1)}  \left( \mathbf{M}  - \lambda_i \mathbf{I}  \right)^{j} \mathbf{G}_i \mathbf{1}   \label{eq-eign-centr-1} \\
\lim_{\kappa \to \frac{1}{\lambda_1}^{-}}  \frac{ \mathbf{g}(\kappa)}{ f(\kappa \lambda_1) \left[ \boldsymbol{\rho}^l \right]^T \mathbf{1}} &=&  \boldsymbol{\rho}^r.  \label{eq:v-limit-lambda}
\end{eqnarray} 
Since $ \left[ \boldsymbol{\rho}^l \right]^T \mathbf{1} >0$, as $\kappa \to \frac{1}{\lambda_1}^{-}$ the rankings produced by 
$\mathbf{g}(\kappa)$ converge to those produced by the entries of $\boldsymbol{\rho}^r$.   

For the behavior as $\kappa \to 0$ we have 
\begin{eqnarray} 
\label{eq:v-limit-0}
\mathbf{g}(\kappa) &=& \mathbf{I}  \mathbf{1} + \kappa \mathbf{M} \mathbf{1} + \kappa^2  \mathbf{M}^2 \mathbf{1} + \ldots  \nonumber \\ 
\frac{\mathbf{g}(\kappa) - \mathbf{1} }{\kappa} & = &  \mathbf{M} \mathbf{1} +  \kappa  \mathbf{M}^2 \mathbf{1} +  \kappa^2  \mathbf{M}^3 \mathbf{1}  \ldots \nonumber \\ 
\lim_{\kappa \to 0^+} \left[ \frac{\mathbf{g}(\kappa) - \mathbf{1} }{\kappa} \right] & = &   \mathbf{M} \mathbf{1}.
\end{eqnarray}
Therefore, it follows that as $\kappa \to 0^+$ the rankings produced by $\mathbf{g}(\kappa)$ converge to those produced by the vector $\mathbf{M} \mathbf{1}$.  

Finally, by substituting $\mathbf{M}$ with $\mathbf{B}$ and $\mathbf{A}^T$, we get the results in Eqs.~(\ref{eq-app-u-1}) and (\ref{eq-app-d-1}). In the special case, when we transform $\kappa \lambda_1 \to \lambda_1$ we are looking at the non-parametrized versions of the OU and ID. Then, Eqs.~(\ref{eq-app-u-1}) and (\ref{eq-app-d-1}) reduce to
\begin{eqnarray}
\mathbf{u} &\approx & 
\left\{
\begin{array}{lll} 
\mathbf{1}  +  \mathbf{B} \mathbf{1}   & \mbox{ for} &  \lambda  \to 0^+,  \\
 \frac{\left[\boldsymbol{\rho}^l(\mathbf{B})\right]^T \mathbf{1}  }{1-\lambda }  \boldsymbol{\rho}^r(\mathbf{B}) & \mbox{ for} & \frac{1}{\lambda}^- \to 1,
\end{array}
\right. \nonumber \\
\mathbf{d} &\approx & 
\left\{
\begin{array}{lll}
\mathbf{1}  +   \mathbf{A}^T \mathbf{1}  & \mbox{ for} & \lambda \to 0^+, \\
\frac{\left[ \boldsymbol{\rho}^r ( \mathbf{A}  )  \right]^T \mathbf{1}}{1-\lambda} \boldsymbol{\rho}^l(\mathbf{A}) & \mbox{ for} & \frac{1}{\lambda}^- \to 1. 
\end{array}
\right. \nonumber 
\end{eqnarray}
where $\lambda = \lambda_1(\mathbf{A}) = \lambda_1(\mathbf{B})$. Summarizing, we have proved that (i) for an economy with $\lambda \to 0$, ranking (ordering) of country-industry pairs depends on out-degree centrality; (ii) for an economy with $\lambda \to 1$, the ranking of sectors depends solely on the network structure.

\section*{Appendix B: Proof of Theorem 2}

Let us study the quasistationary product distribution of the Markov input-chain and the Markov output-chain. This distribution describes the evolution of the state-space of the Markov chain in the regime before the random walker becomes absorbed. Formally for an arbitrary absorbing Markov chain described with a transition matrix $\boldsymbol{\Omega} = \left[ \Omega_{ij} \right]$, the quasistationary product  distribution $\boldsymbol{\pi}$ is defined as
\begin{align*}
    \boldsymbol{\pi} &= \mathbf{\hat{x}} \odot \mathbf{\hat{y}},
\end{align*}
where $\mathbf{\hat{x}}$ and $\mathbf{\hat{y}}$ are correspondingly the right and left eigenvector associated with the largest eigenvalue $\lambda$ of $\boldsymbol{\Omega}$ and $\odot$ is the Hadamard (element-wise) product. The eigenvectors are normalized in a way such that $\sum_i \hat{y}_i = 1$ and $\mathbf{\hat{y}}^T \mathbf{\hat{x}} = 1$.

We claim that the quasistationary product distributions $\boldsymbol{\pi}(\mathbf{A})$ and $\boldsymbol{\pi}(\mathbf{B})$ of the input and output chains are the same. This can be proven in the following way. First, notice that the matrices $\mathbf{A}$ and $\mathbf{B}$ can be written as
\begin{align*}
    \mathbf{A} &= \mathbf{Z} \mathbf{X}^{-1}, \\
    \mathbf{B} &= \mathbf{X}^{-1} \mathbf{Z}, 
\end{align*}
where $\mathbf{X}$ is a diagonal matrix with entries $x_{ij} = x_i$ if $i = j$ and $0$ otherwise. It is widely known that such matrices are similar since we can write $\mathbf{A} = \mathbf{X} \mathbf{B} \mathbf{X}^{-1}$. 

Let $\mathbf{\hat{x}}(\mathbf{A})  = [\hat{x}_1(\mathbf{A}), \ldots, \hat{x}_N(\mathbf{A}) ]^T$ and  $\mathbf{y}(\mathbf{A}  = [\hat{y}_1(\mathbf{A}), \ldots, \hat{y}_N(\mathbf{A})]^T$ be the right and the left dominant eigenvector of the matrix $\mathbf{A}$, respectively. Then. 
\begin{eqnarray*}
\mathbf{A} \mathbf{\hat{x}}(\mathbf{A}) & = & \lambda \mathbf{\hat{x}}(\mathbf{A})  \\
\mathbf{Z} \mathbf{X}^{-1} \mathbf{\hat{x}}(\mathbf{A}) & = & \lambda \mathbf{\hat{x}}(\mathbf{A})  \\ 
\mathbf{X}^{-1} \mathbf{Z} \mathbf{X}^{-1} \mathbf{\hat{x}}(\mathbf{A}) & = & \lambda \mathbf{X}^{-1} \mathbf{\hat{x}}(\mathbf{A})  \\
\mathbf{B} \mathbf{X}^{-1} \mathbf{\hat{x}}(\mathbf{A}) & = & \lambda \mathbf{X}^{-1} \mathbf{\hat{x}}(\mathbf{A})  \\
\mathbf{B} \mathbf{\hat{x}}(\mathbf{B}) & = & \lambda \mathbf{\hat{x}}(\mathbf{B}),
\end{eqnarray*}
where,
\begin{equation} \label{eq-xb}
\mathbf{\hat{x}}(\mathbf{B}) = \mathbf{X}^{-1}\mathbf{\hat{x}}(\mathbf{A}),
\end{equation}
is the right dominant eigenvector of the matrix $\mathbf{B}$. In a similar way, 
\begin{eqnarray*}
\mathbf{\hat{y}}^T(\mathbf{A}) \mathbf{A} & = & \lambda \mathbf{\hat{y}}^T(\mathbf{A}) \\
\mathbf{\hat{y}}^T(\mathbf{A}) \mathbf{X} \mathbf{B} \mathbf{X}^{-1} & = & \lambda \mathbf{\hat{y}}^T(\mathbf{A}) \\ 
\mathbf{\hat{y}}^T(\mathbf{A}) \mathbf{X} \mathbf{B} \mathbf{X}^{-1} \mathbf{X} & = & \lambda \mathbf{\hat{y}}^T(\mathbf{A}) \mathbf{X} \\ 
\mathbf{\hat{y}}^T(\mathbf{A}) \mathbf{X} \mathbf{B} & = & \lambda \mathbf{\hat{y}}^T(\mathbf{A}) \mathbf{X} \\ 
\mathbf{\hat{y}}^T(\mathbf{B}) \mathbf{B} & = & \lambda \mathbf{\hat{y}}^T(\mathbf{B}),
\end{eqnarray*}
where, 
\begin{equation} \label{eq-yb}
\mathbf{\hat{y}}^T(\mathbf{B}) = \mathbf{\hat{y}}(\mathbf{A})^T \mathbf{X},
\end{equation}
is the left dominant eigenvector of the matrix $\mathbf{B}$. Combining equations~(\ref{eq-xb}) and~(\ref{eq-yb}) we obtain 
\begin{eqnarray*}
\mathbf{\hat{x}}(\mathbf{A})  \odot \mathbf{ \hat{y} }(\mathbf{A})  = 
\mathbf{ \hat{x} }(\mathbf{B})  \odot \mathbf{ \hat{y} }(\mathbf{B}),
\end{eqnarray*}
thus concluding the proof.

\section*{Appendix C: Proof of the Theorem 3}

Consider Markov absorbing chain with $J$ absorbing states and a transition matrix: 
\begin{eqnarray*} \label{eq-matrix-Pout-1-1}
{\mathbf{P}_{out}} &=& 
\begin{bmatrix}
\mathbf{I}_{J \times J} & \mathbf{0}_{J \times n} \\
\mathbf{D}_{\eta} & \mathbf{B} 
\end{bmatrix},
\end{eqnarray*}
where $\mathbf{I}_{J \times J}$ is identity matrix,   $\mathbf{0}_{J \times n}$ zero matrix, and $\mathbf{B} = [b_{ij}]$ is the $n\times n$ adjacency output matrix. Let $V = \{1, 2, \ldots, n     \}$ be the set of transition states -- this is the set of all country-industry pairs. Let $S = \{s_1, s_2,  \ldots, s_J     \}$ be the set of all absorbing states.   
Starting in $i$, the process may be absorbed in $s_{\hat{j}} \in S$ in one or more steps. The probability of absorption in a single step is $\eta_{i \hat{j}}$.  If this does not happen, the process may move  either to another absorbing state (in which case it is impossible to reach $s_{\hat{j}}$), or to a transient state $k$. In the latter case there is probability $m_{k \hat{j}}$ of being absorbed in the state $s_{\hat{j}}$). Therefore, we have
$$
m_{i \hat{j}} = \eta_{i \hat{j}} + \sum_{k \in V} b_{ik}  m_{k \hat{j}}
$$
which can be written in matrix form as $\mathbf{M} = \mathbf{D}_\eta + \mathbf{B} \mathbf{M}$. Thus,  
\begin{equation} \label{eq-matrix-M-1}
    \mathbf{M} = ( \mathbf{I} - \mathbf{B}  )^{-1} \mathbf{D}_\eta  
\end{equation}
Note that 
\begin{eqnarray*} 
\lim_{t \to \infty}\left[\mathbf{P}_{out}\right]^t &=& 
\begin{bmatrix}
\mathbf{I}_{J \times J} & \mathbf{0}_{J \times n} \\
\mathbf{M} & \mathbf{0}_{ n \times n} 
\end{bmatrix},
\end{eqnarray*}
hence the matrix $\mathbf{M}$ is row stochastic. Consider now an absorbing Markov chain with one absorbing state and a transition matrix $\mathbf{P}$ given by: 
\begin{eqnarray*} \label{eq-trans-matrix-1}
{\mathbf{P}} &=& 
\begin{bmatrix}
1 & \mathbf{0}^T \\
 \boldsymbol{\alpha}  & \mathbf{Q} 
\end{bmatrix},
\end{eqnarray*}
where $ \boldsymbol{\alpha}$ and $\mathbf{Q}$ are $\boldsymbol{\delta}$ and $\mathbf{A}^T$, respectively,  for the input chain, and $\boldsymbol{\gamma}$ and $\mathbf{B}$, respectively, for the output chain. In this case, it follows from Eq.~(\ref{eq-matrix-M-1}) that 
\begin{eqnarray*}
( \mathbf{I} - \mathbf{A}^T  )^{-1} \boldsymbol{\delta} & = & \mathbf{1} \\
( \mathbf{I} - \mathbf{B}  )^{-1} \boldsymbol{\gamma} & = & \mathbf{1}  
\end{eqnarray*}
Assume now that the row sums of $\mathbf{Q}$  are all equal to a common
value $c<1$. In this case $\boldsymbol{\alpha} = (1-c) \mathbf{1}$. The last two equations can be rewritten as 
\begin{eqnarray*}
( \mathbf{I} - \mathbf{A}^T  )^{-1} \mathbf{1} & = & \frac{1}{1-c}\mathbf{1} \\
( \mathbf{I} - \mathbf{B}  )^{-1} \mathbf{1} & = & \frac{1}{1-c}\mathbf{1},   
\end{eqnarray*}
thus concluding the proof of the theorem.

\vspace{6pt}


\bibliographystyle{unsrt}


\end{document}